\documentclass[preprint,nofootinbib]{revtex4}
\usepackage[dvips]{graphicx}
\usepackage{amsmath,amssymb,axodraw,graphicx,color}

\newcommand{\be}{\begin{equation}}
\newcommand{\ee}{\end{equation}}
\newcommand{\bea}{\begin{eqnarray}}
\newcommand{\eea}{\end{eqnarray}}
\newcommand{\bref}[1]{(\ref{#1})}
\newcommand{\ra}{\rangle}

\newcommand{\la}{\langle}
\newcommand{\bi}{\bibitem}

\begin{document}
\begin{titlepage}
\begin{flushright}
\today
\end{flushright}
\vspace{1cm}
\begin{center}
{\Large\bf Chiral Molecule in the Standard Model.}
\end{center}
\vspace{1cm}
\begin{center}
{\large Takeshi Fukuyama}
\footnote{E-mail:fukuyama@se.ritsumei.ac.jp}
\end{center}
\vspace{0.2cm}
\begin{center}
{\small \it Research Center for Nuclear Physics (RCNP),
Osaka University, Ibaraki, Osaka, 567-0047, Japan}
\end{center}
\vskip 30mm
\begin{center}
{\Large\bf abstract}
\end{center}
This review is based on the talk at the conference of "Spectroscopic Studies on Molecular Chirality" held on Dec 20-21 2013. The objects of the present paper are to (1) derive the 
energy difference between Laevorotatory, or left-handed, (L-) and Dextrotatory, or right-handed, (D-) molecules and to (2) discuss how this tiny energy difference leads us to the observed enantiomer excess. Relations with other parity violating phenomena in molecules, electric dipole moment and natural optical activity, are also discussed.

\end{titlepage}

\section{Standard Model}
We first review the essence of the Standard Model (SM) \cite{Weinberg}, which is necessary for deriving the estimate of the parity violating energy difference in atoms and molecules. SM consists of two ingredients, gauge symmetry and its spontaneous breaking. The gauge principle to construct invariant Lagrangians was first comprehensively discussed by Utiyama \cite{Utiyama}.  Unfortunately, it could not leads us to realsitic weak interaction without spontaneous symmetry breaking mechanism since weak bosons remain massless.  Spontaneous symmetry breaking implies that the ground state is not invariant under the symmetry transformation. The gauge symmetry of the SM is $SU(3)_c\times SU(2)_L\times U(1)_Y$, which is spontaneously broken to $SU(3)_c\times U(1)_{em}$ theory.

The idea of spontaneous symmetry breaking was pioneered by Ginzburg and Landau \cite{GL} as the phenomenological theory of the phase transition of the second kind, 
\be
L=L_0-\frac{{\bf B}^2}{8\pi}-\frac{\hbar^2}{4m}\left|\left(\nabla-\frac{2ie}{\hbar c}{\bf A}\right)\phi \right|^2-a\left|\phi \right|^2-\frac{b}{2}\left|\phi \right|^4.
\ee
Here we have expressed the case corresponding to superconductivity theory of Bardeen-Cooper-Schrieffer \cite{BCS}. $L_0$ is the Lagrangian of the normal state without magnetic field.
This Lagrangian has the minimum if $a<0$
\be
|\phi|^2=-\frac{a}{b}\neq 0.
\label{G-L}
\ee
This gives the skin effect in the London equation.
Nambu-Jona-Lasinio further developed this idea to hadron world, considering the following fermion interaction \cite{N-J}
\be
L=-\overline{\psi}\gamma^\mu\partial_\mu\psi+g\left[\left(\overline{\psi}\psi\right)^2-\left(\overline{\psi}\gamma_5\psi\right)^2\right].
\ee
This Lagrangian is invariant under
\bea
&&\psi\rightarrow e^{i\alpha}\psi,~~\overline{\psi}\rightarrow \overline{\psi}e^{-i\alpha}\label{u1}\\
&&\psi\rightarrow e^{i\alpha\gamma_5}\psi,~~\overline{\psi}\rightarrow \overline{\psi}e^{i\alpha\gamma_5}.\label{chiral}
\eea
If vacuum (expectation value) breaks the chiral symmetry \bref{chiral},
\be
<\overline{\psi}\psi>_0\neq 0,
\label{NJ}
\ee
then hadron has mass
\be
m=-2g<\overline{\psi}\psi>_0.
\ee
Most of mass in the world is due to hadrons and, therefore, to chiral symmetry breaking.

On the other hand, there appears a Nambu-Goldstone boson when the continuous group is broken spontaneosly \cite{N-J}, \cite{NG}. Nicely enough, if this theorem is incorporated in gauge theory, this boson is eaten to the longitudinal part of broken gauge boson and changes it to be massive one. 
Thus weak boson, playing an essential role in this paper, and leptons-quarks have masses by the relativistic version of \bref{G-L} in the minimal coupling of these field with so-called Higgs field \cite{Higgs}.

Combining this symmetry breaking mechanism with the gauge symmetry $SU(3)_c\times SU(2)_L\times U(1)_Y$, the SM was constructed \cite{Weinberg}.

The interaction Lagrangian of leptons with the electro-weak fields in the SM is
\bea
L_{int}^{(e)}=g\left(\overline{L_e}\gamma^\alpha{\bf T}L_e\right)\cdot {\bf A}_\alpha
+g'\left[\textcolor{red}{-\frac{1}{2}}\left(\overline{L_e}\gamma^\alpha L_e\right)+(\textcolor{red}{-1})\left(\overline{R_e}\gamma^\alpha R_e\right)\right]B_\alpha.
\label{ewint}
\eea
Here the first and second terms are $SU(2)_L$ and $U(1)_Y$ gauge interactions, respectively, and ${\bf A}_\alpha~ (g)$ and $B_\alpha~(g')$ are corresponding gauge fields (gauge coupling constants).
The factors $-1/2$ and $-1$ indicate $U(1)_Y$ charges assigned to the SM particles. The quark sector is similarly obtained. See Table I, where left-handed up quark, for instance, is defined by
\be
u_L=\psi^{(u)}_L\equiv \frac{1}{2}\left(1-\gamma_5\right)\psi^{(u)}.
\ee
$u_R^c$ is the charge conjugation of right-handed up quark etc. $h$ is the Higgs doublet.
Color indices for quarks are omitted since they are not concerned with electro-weak interaction.
\begin{table}[h]
\caption{The model is constructed by assigning the following quantum numbers. Quarks and leptons have three families having the same sets of first family explicitly shown in this Table.}
{\begin{tabular}{|c|c|c|c|c|c|}
\hline
 &$SU(3)_c$&$SU(2)_L$&$U(1)_Y$&$U(1)_B$&$U(1)_L$\\
\hline
\hline
$Q_L=\left(
\begin{array}{c}
u\\
d
\end{array}
\right)_L$&$3$&$2$&$+\frac{1}{6}$&$+\frac{1}{3}$&$0$\\
$u_R^c$&$3^*$&$1$&$-\frac{2}{3}$&$-\frac{1}{3}$&$0$\\
$d_R^c$&$3^*$&$1$&$+\frac{1}{3}$&$-\frac{1}{3}$&$0$\\
$L_e=\left(
\begin{array}{c}
\nu_e\\
e
\end{array}
\right)_L$&$1$&$2$&$-\frac{1}{2}$&$0$&$+1$\\
$e_R^c$&$1$&$1$&$+1$&$0$&$-1$\\
$h=\left(
\begin{array}{c}
h^0\\
h^-
\end{array}
\right)$&$1$&$2$&$-\frac{1}{2}$&$0$&$0$\\
\hline
\hline
\end{tabular}}
\end{table}

After the symmetry breaking of $SU(2)_L\times U(1)_Y$ to $U(1)_{em}$, the third (isospin) component of gauge fields ${\bf A}_\alpha$ and $B_\alpha$ are mixed to give rise to the electromagnetic and neutral weak fields,
\bea 
A_\mu&=&\mbox{cos}\theta B_\mu+\mbox{sin}\theta A_\mu^3~~\mbox{: electromagnetic field}, \nonumber\\
Z_\mu&=&-\mbox{sin}\theta B_\mu+\mbox{cos}\theta A_\mu^3~~\mbox{: neutral weak boson}.
\label{mixing}
\eea
Here $\theta$ is the Weinberg angle and experimentally determined as sin$^2\theta=0.23$.
Substituting \bref{mixing} into \bref{ewint}, we obtain 
\bea\label{neutral}
&&L_{int}=\frac{g}{\sqrt{2}}\overline{L}_e\gamma^\alpha (T_-W_\alpha^-+T_+W_\alpha^+)L_e\nonumber\\
&&+\left[g\cos \theta\overline{L}_e\gamma^\alpha T_3L_e+g'\sin\theta\left(\frac{1}{2}\overline{L}_e\gamma^\alpha L_e+\overline{R}_e\gamma^\alpha R_e
\right)\right]Z_\alpha\\
&&+\left[\textcolor{red}{-g'\cos\theta}\left(\frac{1}{2}\overline{L}_e\gamma^\alpha L_e+\overline{R}_e\gamma^\alpha R_e\right)+\textcolor{red}{g\sin\theta}\overline{L}_e\gamma^\alpha T_3L_e\right]A_\alpha\nonumber\\
&&\equiv J_-^{(e)\alpha}W_\alpha^-+J_+^{(e)\alpha}W_\alpha^++J_0^{(e)\alpha}Z_\alpha-eJ_{EM}^{(e)\alpha}A_\alpha.
\eea
$A_\alpha$ is the electromagnetic field, and $e=g\sin\theta=g'\cos\theta$ or equivalently $e=\frac{gg'}{\sqrt{g^2+g'^2}}.$
The second term of \bref{neutral} is the neutral current which takes essential roles in this paper,
\bea
L_{neutral}=\left[g\cos \theta\overline{L}_e\gamma^\alpha T_3L_e+g'\sin\theta\left(\frac{1}{2}\overline{L}_e\gamma^\alpha L_e+\overline{R}_e\gamma^\alpha R_e
\right)\right]Z_\alpha.\nonumber
\eea

\section{How does L- and D-molecular energy difference appear ?}
Assigning the $U(1)_Y$ charge of quarks in Table I, we can easily generalize the neutral current to lepton-quark systems,
\be
H_{int}=\left(J_0^{(e)\mu}(x)+J_0^{(u)\mu}(x)+J_0^{(d)\mu}(x)\right)Z_\mu.
\ee
Here $J_0^{(e)\mu},~J_0^{(u)\mu}$ and $J_0^{(d)\mu}$ are the neutral currents of electron, up quark, and down quark, respectively.
Their explicit forms are 
\be
J_0^{(i)\mu}(x)\equiv \frac{g}{4\cos\theta}\epsilon_i\overline{\psi}_i(x)\gamma^\mu(C_V^i-\gamma_5)\psi_i(x),~~(i=e,u,d)
\ee
with
\bea
&&\epsilon_u=1,~\epsilon_d=\epsilon_e=-1,\nonumber\\
&&C_V^e=1-4\sin^2\theta,~~C_V^u=1-\frac{8}{3}\sin^2\theta,\\
&&C_V^d=1-\frac{4}{3}\sin^2\theta.\nonumber
\eea

Proton and neutron are composed of $p=uud$ and $n=udd$, and
\be
J_\mu^p=2J_\mu^u+J_\mu^d,~~J_\mu^n=J_\mu^u+2J_\mu^d.
\ee
Parity violating electron-nucleon interaction is obtained by contracting neutral weak boson $Z$ which intermediates electron and nucleon currents with mass $M_Z$,
\bea
H_{eN}^{PV}&=&-\frac{G_F}{\sqrt{2}}(e^\dagger (x)\gamma_5 e(x))\left(\frac{Z-N}{2}-2\mbox{sin}^2\theta Z\right)\\
&\equiv&-\frac{G_F}{\sqrt{2}}(e^\dagger (x)\gamma_5 e(x))\frac{Q_W}{2}. \nonumber
\eea
The Fermi constant $G_F$ is defined by  $G_F/\sqrt{2}=g^2/(8M_Z^2\cos^2\theta)$ and $Q_W$ is called weak charge.
In the nonrelativistic limit, electron wave function is described as
\bea
e(x)=\psi(x)=\left(
\begin{array}{c}
\varphi(x)\\
\frac{{\bf \sigma}\cdot {\bf p}}{2m_e}\varphi(x)
\end{array}
\right)
\eea
and
\bea
\langle a|e^\dagger (x)\gamma_5 e(x)|b\rangle=\psi_a^*(x)\gamma_5\psi_b(x)
=\left[\varphi_a^*\left(\frac{{\bf \sigma}\cdot{\bf p}}{2m_e}\varphi_b\right)+\left(\frac{{\bf \sigma}\cdot{\bf p}}{2m_e}\varphi_a\right)^*\varphi_b\right].
\eea
Then we obtain parity-violating potential becomes
\bea
V_{eN}^{PV}=\frac{G_F\alpha^2}{4\sqrt{2}}\sum_{a,i}Q_W^a\{{\bf\sigma}_i\cdot{\bf p}_i,\delta^3({\bf r}_{ia})\}_+~~\mbox{in atomic units (a.u.),}
\label{eN}
\eea
where ${\bf r}_{ia}$ is the distance from $i$'th electron to the center of $a$'th nucleus.
Let us consider Z-scaling for $V_{PV}$ \cite{Zeldovich}\cite{Bouchiat}. In terms of the radial wave function $R_{nl}$. the matrix element of $V_{PV}^{eN}$ becomes
\bea
\langle ns_{1/2}|V_{PV}^{eN}|n'p_{1/2}\rangle&=&\frac{3i}{16\pi m_ec}\left(\frac{G_F}{\sqrt{2}}\right)Q_W(Z,N)R_{n0}(0)\frac{dR_{n'1}(0)}{dr}\nonumber\\
&\approx & i\frac{\hbar (G_F/\sqrt{2})Q_WZ^2}{4\pi m_eca_B^4(\nu_n\nu_{n'})^{3/2}}
\eea
with the effective principal quantum number $\nu_n$ and the Bohr radius $a_B$.
Here use has been made of 
\be
R_{n0}\approx \frac{2Z^{1/2}}{a_B^{3/2}\nu_n^{3/2}},~~\frac{dR_{n'1}(0)}{dr}\approx \frac{2}{3}\frac{Z^{3/2}}{a_B^{5/2}\nu_n^{3/2}}.
\ee
Electron-electron parity-violating potential, on the other hand, is given by
\bea
V_{ee}^{PV}=-\frac{G_F\alpha}{2\sqrt{2}}(1-4\mbox{sin}^2\theta)\{{\bf \sigma}\cdot{\bf p},~n_e(x)\}_+
\eea
for a single valence electron case, and 
\bea
r\equiv |\frac{\la ns|V_{ee}^{PV}|n'p\ra}{\la ns|V_{eN}^{PV}|n'p\ra}|<\frac{112}{9\pi}\frac{1-4\sin^2\theta}{(1-4\sin^2\theta)Z-N}.
\eea
Taking $\sin^2\theta=0.23,~Z=55,~N=78,$ we obtain $r<4.3\times 10^{-3}$ \cite{Bouchiat}.
So hereafter we consider only e-N interaction for PV interaction.
In nonrelativistic limit, the molecular wave function may always be chosen to be real, while the coordinate part of $H^{PV}$ is pure imaginary. So the expectation value of $H^{PV}$ is zero.
In order to get non-zero value we must invoke spin-orbit interaction,
\be
V^{SO}=\frac{e\hbar}{4m_e^2c^2}({\bf E}\times {\bf p}_a)\cdot{\bf \sigma}_a=\frac{\hbar^2}{2m_e^2c^2r}\frac{dU}{dr}{\bf l}_a\cdot {\bf s}_a\equiv f(r)_a{\bf l}_a\cdot {\bf s}_a.
\label{SO}
\ee
Here we have used ${\bf E}=\frac{{\bf r}}{r}\frac{dU}{dr}$. Thus the second-order perturbation is given by the replacement,
\bea
\psi_0\rightarrow \psi_0+\sum_n\frac{\langle n |V^{SO}|0\rangle}{E_0-E_n}\psi_n\nonumber
\eea
and
\bea
&&\Delta E^{PV}=\sum_n\frac{\la 0|V^{PV}|n\ra\la n|V^{SO}|0\ra}{E_0-E_n}+c.c.\nonumber\\
&&=\frac{G_F}{\sqrt{2}}\sum_{n,\alpha,\beta}Q_W\frac{\la \varphi_i|\{{\bf p},~\delta^3({\bf r}_\alpha)\}_+|\varphi_n\ra\cdot\la \varphi_n|f({\bf r}_\beta){\bf l}_\beta |\varphi_i\ra}{\epsilon_i-\epsilon_n}.
\label{DeltaEPV}
\eea
The order estimation of $V^{SO}$ is as follows.
The main contribution is given by distance close to the nucleus
  of the order of the Bohr radius, $\hbar^2/(Zm_ee^2)\equiv a_B/Z$, and
\be
|U(r)|\approx Ze^2/r\rightarrow Z^2
\ee
in a.u. and
\be
f\approx \hbar^2U/(m_e^2c^2r^2)\rightarrow Z^4\alpha^2.
\ee
The mean value of $f$ is 
\be
\overline{f}=fw\approx fZ^{-2}\approx Z^2\alpha^2.
\ee
Here $w$ is the probability of finding the electron in the region $r\leq 1/Z$ and given by \cite{Landau}
\be
w\approx |\psi|^2r^3\approx Z^{-2},
\ee
where use has been made of
\be
|\psi|\approx \frac{1}{r\sqrt{p}}\approx \frac{1}{r|U|^{1/4}}\approx \sqrt{Z}.
\ee
So if $E_0-E\approx 1$ eV, then we obtain
\bea
\Delta E^{PV}\approx G_F\alpha^4Z^5\approx 8.5\times 10^{-21}Z^5~ \mbox{a.u.}=2.3\times 10^{-19}Z^5 ~\mbox{eV},
\eea
where we have set $Q_W\approx -N\approx -Z$.\footnote{N dependence may be important for the isotopic effect of the abundances of carbon ${}^{12}C$ and ${}^{13}C$ in the biological molecules where ${}^{13}C$ is less abundant than in nature.}

So far we have not discussed molecules explicitly. Chiral property concerning with this paper appears first when we consider molecules. The energy difference of chiral molecules is schematically described in Fig.1. It should be remarked that $\Delta E_{PV}^*$ and $\Delta E_{PV}$ must be different otherwise no difference appears in L- and D-molecules. 
\begin{figure}[h]
\begin{center}
\includegraphics[width=0.49\textwidth]{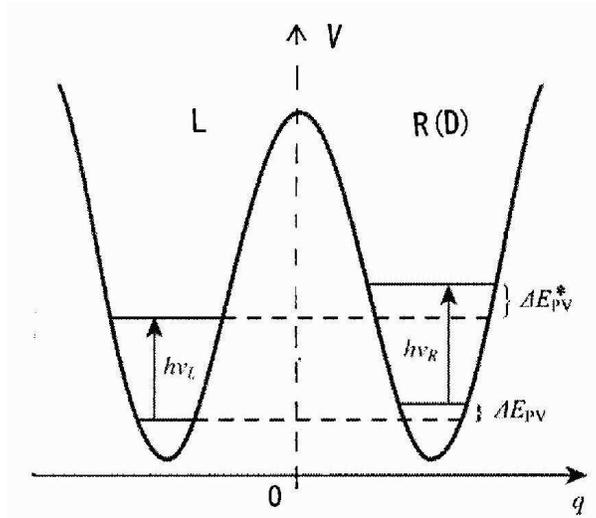}
\caption{The energy difference between L- and D-molecules. $q$ is a spatial coordinate perpendicular to mirror. It should be remarked that $\Delta E_{PV}^*$ and $\Delta E_{PV}$ must be different otherwise no difference appears in L- and D-molecules.}
 \label{fig:mp}
\end{center}
\end{figure}
The expectation value of spin orbit interaction of alkali atom \bref{SO} is
\be
V^{SO}=f(r)\left(J(J+1)-L(L+1)-S(S+1)\right).
\label{SO}
\ee
For the transition from $S_{1/2}$ to $P_{1/2}$, $V^{SO}$ gives different energy shift at each level and, therefore, different $\Delta E^{PV}$.

If we adopt the familiar Linear Combination of Atomic Orbitals (LCAO) approximation, $\Delta E^{PV}$ of \bref{DeltaEPV} is modified for molecules as \cite{Sandars}
\be
\Delta E^{PV}=\frac{G_F}{\sqrt{2}}\sum_{n,\alpha,\beta} c_{i\gamma}^\alpha c_{n\gamma'}^\alpha c_{n\gamma''}^\beta c_{i\gamma'''}^\beta Q_W^\alpha\frac{\la \varphi_\gamma^\alpha|\{{\bf p},~\delta^3({\bf r}_\alpha)\}_+|\varphi_{\gamma'}^\alpha\ra\cdot\la \varphi_{\gamma''}^\beta|f({\bf r}_\beta){\bf l}_\beta |\varphi_{\gamma'''}^\beta\ra}{\epsilon_i-\epsilon_n}.
\ee
Here $i~(n)$ indicates initial (intermediate) state, and molecular wave function is expanded as
\be
\psi=\sum c_\gamma^\alpha \varphi_\gamma^\alpha,
\ee
where $\varphi_\gamma^\alpha$ are atomic orbitals centered on nuclei $\alpha$.  

For more detailed calculation, we need complicated calculations of wave function in general. Parity violation indeed appears in many places in molecules other than the electron term.
In the case of electric dipole moments polar molecules take very important roles since small (smaller than electron terms by factor $m_e/M_{nucleus}$) nuclear rotation energy level is used to induce very huge internal electric field, leading to huge molecular EDM \cite{Fukuyama}. We will discuss parity violation in molecule in more detail in Discussions.
For chiral molecules, the selection rules due to $\delta({\bf r})$ term and the spin-axis interaction of \bref{DeltaEPV} give some restriction. 

Parity-violating energy difference in molecules is evaluated \cite{Sandars} \cite{Quack} and
\be
\Delta E^{PV}\approx 4f_{geo}\times 10^{-21}Z_{eff}^5~ \mbox{eV}, 
\ee
where $f_{geo}$ is a geometry-dependent empirical factor and usually smaller than $1$.
 However, as we mentioned, many points are left ambiguous in molecule. Detailed arguments on molecules are out of the scope of this review and will be discussed in a separate form.

\section{How does tiny L-D energy difference cause the observed enantiomer excess ?}
We have known that the Standard Model produces the L-D energy difference
but it is so tiny. So we need its globarization even if it is the origin of the enantiomer excess of our world.
Here we consider linear amplification model \cite{Yamagata} and nonlinear (an auto-catalytic) model \cite{Kondepudi} as illustrations. 
\subsection{linear amplification model}
Let us consider D- and L- molecule $A$ and $A'$ which are polymerized to D- and L-type deoxyribonucleic acid (DNA) by many steps,
\bea 
A_1\rightarrow A_2\rightarrow A_3\rightarrow ....\mbox{real (D-) DNA},\\
A'_1\rightarrow A'_2\rightarrow A'_3\rightarrow ....\mbox{imaginary (L-) DNA}.
\eea
Let us denote the ratio of reaction rate of $k$th reaction in the real series
relative to imaginary series by $p_k$.
Then a final ratio of final products will be 
\be
\frac{N_{real}}{N_{imag}}=p_1p_2...p_n.
\ee
For simplicity we assume all $p_k=p=1+g,~g=\Delta E^{PV}/kT\ll 1$.  
\be
\frac{N_{real}}{N_{imag}}=p^n=(1+g)^n\approx e^{gn}.
\ee
$n$ may be of order of the number of nucleotides in a cell $1\times (10^8-10^9)$. Thus even if $g\ll 1$, the observed enantiomer excess may be realized. 

\subsection{auto-catalytic model}
Next we explain the other nonlinear scenario, that is, auto-catalytic process \cite{Kondepudi}. 
We start with matters A,B which have no chirality, making the following reactions:

(i) A and B combine to produce L-handed molecule $X_L$ and D-handed molecule $X_D$:
\bea
&&A+B\stackrel{K_{1L}}{\rightarrow}X_L,~~A+B\stackrel{K_{-1L}}{\leftarrow}X_L,\\
&&A+B\stackrel{K_{1D}}{\rightarrow}X_D,~~A+B\stackrel{K_{-1D}}{\leftarrow}X_D,\nonumber
\eea
where $K_{\pm 1L}$ and $K_{\pm 1D}$ are the corresponding reaction rates.

(ii) $X_L$ and $X_D$ can autocatalitically reproduced:
\bea
&&X_L+A+B\stackrel{K_{2L}}{\to}2X_L,~~X_L+A+B\stackrel{K_{-2L}}{\leftarrow}2X_L,\\
&&X_D+A+B\stackrel{K_{2D}}{\to}2X_D,~~X_D+A+B\stackrel{K_{-2D}}{\leftarrow}2X_D.\nonumber
\eea
(iii) $X_L$ and $X_D$ react to form achiral D irreversibly,
\be
X_L+X_D\stackrel{K_3}{\to}D.
\ee
The system is assumed to be open to maintain the concentration of A,B 
constant and D is removed continually and back reaction 
of (iii) is eliminated. Thus the kinetic equations of this 
system become
\bea\label{kinetic1}
\frac{dX_L}{dt}&=&K_{1L}(AB)-K_{-1L}(X_L)+K_{2L}(ABX_L)-K_{-2L}(X_L^2)-K_3(X_LX_D),\\
\frac{dX_D}{dt}&=&K_{1D}(AB)-K_{-1D}(X_D)+K_{2D}(ABX_D)-K_{-2D}(X_D^2)-K_3(X_LX_D),
\label{kinetic2}
\eea
where $(AB)$ etc. are the concentrations of the corresponding AB etc. 
We will discuss the intermediate states $X_L^*,~X_D^*$ with energy difference
$\Delta E$, and
\bea
\frac{K_L}{K_D}=e^{\Delta E/kT}\simeq 1+(\Delta E/kT)\equiv 1+g.
\label{DeltaE}
\eea
Before doing that, we first study simpler case where $K_{\pm iL}=K_{\pm iD}$ (i=1,2).
This is the case when there is no parity-violating interaction.
Let us consider the steady states, $dX_L/dt=dX_D/dt=0$. If $(AB)>(AB)_c$, the symmetric state becomes unstable and new symmetry-breaking state becomes possible.
To express this, let's introduce
\bea
\alpha\equiv ((X_L)-(X_D))/2,~~ \beta\equiv ((X_L)+(X_D))/2.
\eea
Then \bref{kinetic1} and \bref{kinetic2} are rewritten as
\bea\label{kinetic3}
\frac{d\alpha}{dt}&=&\{-K_{-1}+K_2(AB)-2K_{-2}\beta\}\alpha ,\\
\frac{d\beta}{dt}&=&K_1(AB)-K_{-1}\beta+K_2(AB)\beta-K_{-2}(\beta^2+\alpha^2)-K_3(\beta^2-\alpha^2).
\label{kinetic4}
\eea

If $(AB)<(AB)_c$ the symmetric steady solution appears:
\bea
&&\alpha_S=0,\\
&&\beta_S=\frac{2K_{-2}\beta_A+\left[(2K_{-2}\beta_A)^2+4(K_{-2}+K_3)K_1AB\right]^{1/2}}{2(K_{-2}+K_3)},
\eea
where $\beta_A$ is given by \bref{betaA}.

If $(AB)$ exceeds $(AB)_c$, the system will be driven to one of the asymmetric state.
\bea
\label{alphaA}
\alpha_A&=&\pm\left(\frac{\beta_A^2-K_1(AB)}{K_3-K_{-2}}\right)^{1/2},\\
\beta_A&=& (K_2AB-K_{-1})/2K_{-2}.
\label{betaA}
\eea
$\alpha$ versus $(AB)$ relation is described in Fig.2(a). Thus even if parity is not violated,
symmetric phase is not stable. However, $\pm$ asymmetric phases can be equally produced.
Neither L- nor D-molecule is predominant, of course.
\begin{figure}[h]
\begin{center}
\includegraphics[width=0.49\textwidth]{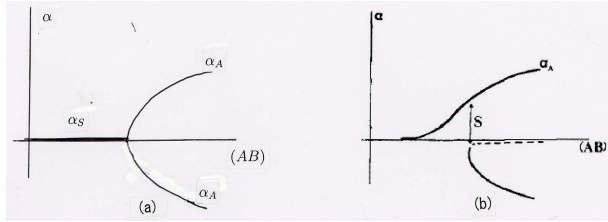}
\caption{The symmetric phase becomes unstable if $(AB)$ exceeds $(AB)_c$. (a) and (b) are the cases in the absence and presence of parity-violating interaction, respectively. In the latter case L- and D-molecule bifurcate. S is proportional to $g^{1/3}$ (See \bref{S}). Figures are cited from \cite{Kondepudi}.}
 \label{fig:bifur}
\end{center}
\end{figure}

If we give asymmetric initial condition between L- and D-molecules concentrations and if there is strong autocatalytic reaction, then there appears non zero $\alpha_A$ phase even if there is no reaction rate difference between L- and D-molecules (Fig.3).
\begin{figure}[h]
\begin{center}
\includegraphics[width=0.29\textwidth]{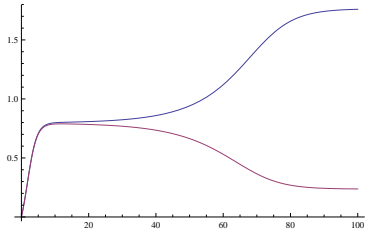}
\caption{The growth of asymmetric autocatalysis due to tiny difference of the initial conditions $(X_L(0))=0.002$ and $(X_D(0))=0.000$ with $(A)=0.5, ~(B)=0.5~K_{-1}=0.1,~K_2=2.0,~K_{-2}=0.2,~K_3=0.5.$ Horizontal axis is time. The upper (lower) line is $(X_L)~ ((X_D))$. The splitting ($\alpha\neq 0$) occurs when the auto catalytic process is strong (large $K_2$).} 
 \label{fig:kinetic}
\end{center}
\end{figure}

Remark that the asymmetric phase appears when
$K_2$ is much larger than the other $K$'s unlike the book of Prigogine-Kondepudi \cite{KP}.
However, in this case we have set the initial condition by hand.

In the presence of $\Delta E^{PV}$, we set
\be
K_{1L}=(1+g/2)K_1,~~K_{1D}=(1-g/2)K_1
\ee
and ignore the effect of the perturbation on the other kinetic constants and $K_{-1L}=K_{-1D}$ etc.
\bea
\frac{d\alpha}{dt}&=&-U\alpha^3+V((AB)-(AB)_c)\alpha+Wg\\
&=&-U(\alpha-\alpha_0)^2(\alpha-\alpha_1).\nonumber
\eea
Here
\bea 
&&U=\frac{2K_{-2}^2(K_3-K_{-2})}{K_3(K_2(AB)_c-K_{-1})},\\
&&V=K_2-\frac{K_{-2}}{K_3}\left(K_2+\frac{2K_1K_{-2}}{K_2(AB)_c-K_{-1}}\right),\\
&&W=\frac{1}{2}K_1(AB)_c.
\eea

Then one of $\alpha_A$ phases bifurcates from the other one depicted in Fig.2(b).
$S$ is given by
\be
S=\alpha_1-\alpha_0=3\left(\frac{Wg}{2U}\right)^{1/3}.
\label{S}
\ee
\section{Discussions}
We have discussed in this review how the energy difference between L- and D-molecules in the framework of the standard model. The difference is very tiny but some grow-up mechanisms have been briefly discussed. Unfortunately the seed of such mechanism does not necessarily depend on the energy difference due to the parity violating neutral current.
Z scale dependence of the energy difference becomes less clear in molecular case than in atomic case and need further study. 
Here we briefly discuss the physical implications of space-time symmetry breaking in molecule. 
Let us consider three cases, electric dipole moment (EDM) of molecule and natural optical activity together with the chiral molecoe (\bref{DeltaEPV}). They have several common properties: they break parity symmetry and appear as relativistic and finite size effects. EDM of paramagnetic atom (and molecule) is described as
\be
{\bf d}_{atom}=\sum_n\frac{\left<0|e\sum_i^Z{\bf r}_i|n\right>\left<n|\sum_i^Z(\gamma_{0,i}-1)\Sigma_i\nabla_i\Phi ({\bf r})|0\right>}{E_0-E_n}+h.c.
\label{para},
\ee
where
\begin{eqnarray}
\boldsymbol{\Sigma}\equiv \left(
   \begin{array}{cc}
    \boldsymbol{\sigma} & {\bf 0}\\
    {\bf 0} & \boldsymbol{\sigma}
   \end{array}
   \right).
\label{Sigma}
\end{eqnarray}

On the other hand, natural optical activity is given by \cite{Condon}
\be
\beta_i=\frac{2c}{3\hbar}\sum_n\frac{\la i|e\sum_i{\bf r}_i|n\ra\cdot\la n| {\bf \mu}|i\ra}{\omega_{ni}^2-\omega^2}.
\label{natural}
\ee
Here $\beta$ is related with complex refractive index
\be
n_{\pm}=\epsilon^{1/2}\pm 2\pi g\nu
\ee
with $4\pi N\beta/(3c)=g/(\epsilon +2)$. $N$ is the number of molecules in unit volume and
\be
{\bf \mu}=\frac{e}{2mc}\sum_i\left({\bf r}_i\times{\bf p}_i+2{\bf s}_i\right).
\ee

$g$ comes from the finite size effect of dielectric constant \cite{L-L},
\be
\epsilon_{ik}(\omega,{\bf k})=\epsilon_{ik}^{(0)}(\omega)+i\gamma_{ikl}k_l=\epsilon_{ik}^{(0)}(\omega)+i\epsilon_{ikm}g_{ml}n_l.
\ee
For an opticall active isotropic body, $g_{ml}=f\delta_{ml}$ and we obtain double circular refraction,
\be
n_\pm^2=n_0^2\pm g,
\ee
where $n_0=\sqrt{\epsilon^{(0)}}$ and $g=fn_0$.
The difference of energy dependence in \bref{natural} from the others is due to periodic condition of perturbation of light wave and reduces to the analogous depence to the others for $\omega=0$. Thus \bref{natural} seems to be situated in the intermediate stage between EDM (\bref{para}) and the chiral difference (\bref{DeltaEPV}) via $e{\bf r}_i$ and ${\bf l}_i$.

Though very briefly, we have discussed the origin and the growth of enantiomer. Problem how to detect the existing enantiomer excess is another important problem, on which we must cite two big works presented in this workshop:
Hirota gave a nice method to detect enantiomer difference using three types of rotational spectra \cite{Hirota} and Doyle et al. realized this method experimentally in a very beautiful way \cite{Doyle}. Finally we comment that in the recent development of molecule spectroscopy, direct detection of $\Delta E^{PV}$ becomes the target of on-going experiments \cite{Darquie}.
\subsection*{Acknowledgements}
We are grateful to Dr.Momose for inviting the author to the workshop of Spectroscopic Structures on Molecular Chirality. We also thank Drs. M.Quack, H.Kanamori, T.Shida, T.Aoki, and H.Sugiyama for useful discussions.

\end{document}